\documentclass[11pt]{article}
\usepackage[normalem]{ulem}
\usepackage{jheppub}
\usepackage{color}
\usepackage[table]{xcolor}
\usepackage{epstopdf}
\usepackage{cleveref}
\usepackage{epstopdf}
\usepackage{caption}
\usepackage{subcaption}
\usepackage{amsthm}
\usepackage{slashed}
\usepackage{orcidlink}

\DeclareMathOperator*{\sgn}{sgn}
\DeclareMathOperator{\cut}{Cut}

\newcommand{\pol}{\ensuremath{\varepsilon}}
\newcommand{\ctpeps}{\ensuremath{i0}}
\newcommand{\dimreg}{\ensuremath{\epsilon}}
\newcommand{\odimreg}[1]{\ensuremath{\mathcal{O}(\dimreg^{#1})}}

\newcommand{\msggs}{\ensuremath{\mathcal{M}_{sggs}}}

\newcommand{\graph}[1]{G_{\text{#1}}}
\newcommand{\rcut}[1]{\ensuremath{\overline{\cut}_{#1}}}

\newcommand*\diff{\mathop{}\!\mathrm{d}}
\newcommand*\Diff[1]{\mathop{}\!\mathrm{d}^#1}

\newcommand{\intw}{\ensuremath{\int \frac{\diff \omega}{2 \pi}}}
\newcommand{\inte}[1]{\ensuremath{\int_{-\infty}^{\infty} \diff \omega \kappa(\omega) \omega^{#1}}}
\newcommand{\intem}[1]{\ensuremath{\int_{-\infty}^{\infty} \diff \omega \kappa(\omega,\mu) \omega^{#1}}}

\newcommand{\deltaE}[2]{\ensuremath{(\Delta E)^{\text{#1}}_{\text{#2}}}}
\newcommand{\delEInc}[1]{\ensuremath{(\overline{\Delta E})_{\text{#1}}}}

\newcommand{\lpegm}{\ensuremath{\log\left(\frac{\omega^2 e^{\gamma_E}}{\mu^2\pi}\right)}}

\newcommand{\partel}[1]{\ensuremath{x_{#1}}}

\newcommand{\cutcoef}[1]{\ensuremath{\mathcal{T}_{#1}}}
\newcommand{\cutcoefr}[1]{\hyperref[eq:ccoef-#1]{\cutcoef{#1}}}

\newcommand{\eg}{{e.g.~}}
\newcommand{\ie}{{i.e.~}}

\colorlet{cutcolor}{blue!60!white}
\colorlet{blobcolor}{gray!40}

\colorlet{ref-link}{green!70!black!70!blue}
\colorlet{cite-link}{red!60!blue}

\colorlet{fc}{green!50!black}
\colorlet{alex}{blue!50!green}

\hypersetup{
  citecolor = cite-link,
  linkcolor = ref-link,
}

\newcommand{\inctr}{\href{https://gitlab.com/aedison/increasingtrees}{IncreasingTrees}}

\newcommand{\prevpaper}{Ref.~\cite{Edison:2023qvg}}

\usepackage{tikz}
\usetikzlibrary{calc}
\usetikzlibrary{patterns}
\usetikzlibrary{decorations.pathreplacing}
\usetikzlibrary{decorations.markings}
\usetikzlibrary{decorations.pathmorphing}
\usetikzlibrary{positioning}
\usetikzlibrary{bending}

\pgfdeclarelayer{background}
\pgfsetlayers{background,main}

\usetikzlibrary{shapes,shapes.geometric,arrows,arrows.meta,decorations.pathmorphing,decorations.markings,decorations.pathreplacing,patterns}
\usetikzlibrary{tikzmark}

\tikzset{every picture/.style={baseline={([yshift=-.7ex]current bounding box.center)}}}

\tikzset{every node/.style={font=\scriptsize}}

\tikzset{dir/.style={decoration={markings, mark=at position \halfway with {\arrow{Latex}}},postaction={decorate}}}
\tikzset{cuts/.style={dash pattern=on 4pt off 2pt,draw=cutcolor,ultra thick}}

\tikzset{photon/.style={decorate,decoration={coil,aspect=0,segment length=2,amplitude=1}}}
\tikzset{photon big/.style={decorate,decoration={coil,aspect=0,segment length=4,amplitude=2}}}

\tikzset{potential/.style={%
   photon big}}

\tikzset{potentialf/.style={%
    dashed,thick,gray}}

\tikzset{quad src/.style n args={2}{%
    fill,anchor=center,inner sep=#1,minimum size=#1,label=left:{#2}}}
\tikzset{mass src/.style n args={1}{%
    circle,fill,anchor=center,minimum size=#1,inner sep=#1,label=left:{\(E\)}}}
\tikzset{blob/.style n args={1}{%
    draw=black,fill=blobcolor,circle,minimum size=#1,inner sep=0}}
  
\tikzset{bulk/.style={%
    circle,fill=blobcolor,draw=black,inner sep=1pt}}

\tikzset{pics/cut hill/.style n args={2}{code={%
      \draw[photon big] (#1.center)
      .. controls ($(#1) + (0.7,0)$) and ($(#2)+(0.7,0)$)..
        node[pos=1/2](v){} (#2.center);
        \draw[cuts] ($(v)+(right:0.3)$) -- ++(left:0.6);
      }}}

\tikzset{pics/cut vhill/.style n args={2}{code={%
      \draw[photon big] (#1.center)
      .. controls ($(#1) + (0,0.7)$) and ($(#2)+(0,0.7)$)..
        node[pos=1/2](v){} (#2.center);
        \draw[cuts] ($(v)+(up:0.3)$) -- ++(down:0.6);
      }}}
\tikzset{pics/cut dhill/.style n args={2}{code={%
      \draw[photon big] (#1.center)
      .. controls ($(#1) + (0,-0.7)$) and ($(#2)+(0,-0.7)$)..
        node[pos=1/2](v){} (#2.center);
        \draw[cuts] ($(v)+(up:0.3)$) -- ++(down:0.6);
        }}}

\tikzset{pics/hills/.style args={#1}{code={%
      \node[quad src={3}{\(I^{ij}\)}] (0) at (0,0){};
      \node[quad src={3}{\(I^{mn}\)}] (b) at (0,#1+1){};
      \foreach \ml[remember=\ml as \lastm] in {1,...,#1}
      {
        \node[mass src={2}](\ml) at (0,\ml) {};
        \draw pic {cut hill={\lastm}{\ml}};
        }
        \draw pic {cut hill={#1}{b}};
    }}}

\pgfmathsetmacro{\mampPushback}{1.2}
\tikzset{
  mampControlsCoords/.style={
    execute at begin picture={
      \coordinate (mampCC) at (\mampPushback,0);
   }
 },
every picture/.append style={mampControlsCoords} 
}

\tikzset{pics/mamp/.style n args={3}{code={%
      \pgfmathsetmacro\n{int(#3+2)}
      \draw[photon big](#1.center)
      .. controls ($(#1) + (mampCC)$) and ($(#2)+(mampCC)$)..
      node[bulk,pos=1/2](v){$\mathcal{M}_{\n}$} (#2.center);
      \foreach \ml in {1,...,#3}
      {
        \node[mass src={2}](\ml) at ($(#1)+(up:\ml)$){};
        \begin{pgfonlayer}{background}
          \draw[potential](\ml) -- (v.center);
        \end{pgfonlayer}
      }
      \draw[cuts] ($(v) +(1,0)+(220-90:1.5)$) arc (220-90:320-90:1.5);
    }}}

\tikzset{pics/split/.style n args={4}{code={%
      \pgfmathsetmacro\nt{int(#2+#3+2)}
      \pgfmathsetmacro\nm{int(#2+#3)}
      
      \pgfmathsetmacro\mno{int(#2+2)}
      \pgfmathsetmacro\mnt{int(#3+2)}
      \pgfmathsetmacro\pf{1/3}
      \pgfmathsetmacro\ps{2/3}

      \draw[photon big](#1) to[out=10,in=-10]
      node[bulk,pos=\pf] (v1){$\mathcal{M}_{\mno}$}
      node[bulk,pos=\ps] (v2){$\mathcal{M}_{\mnt}$} (#4) ;

      \node (cc) at ($(v1)!0.5!(v2)$){};
      \node (ci) at ($(cc)+(180:0.7)$) {};
      \node (cf) at ($(cc)+(0:0.7)$) {};

      \begin{pgfonlayer}{background}
        \foreach \ml in {1,...,#2}
        {
          \node[mass src={2}] (\ml) at ($(#1)+(0,\ml)$){};
          \draw[potential] (\ml) -- (v1.center);
        }
        
        \pgfmathsetmacro\ss{int(#2+1)}
        \foreach \ml in {\ss,...,\nm}
        {
          \node[mass src={2}] (\ml) at ($(#1)+(0,\ml)$){};
          \draw[potential] (\ml) -- (v2.center);
        }
      \end{pgfonlayer}
      \draw[cuts] (ci)--(cf);
      \draw[cuts] ($(v1) +(1,0)+(145:1.7)$) arc (145:360-135:1.7);
      \draw[cuts] ($(v2) +(1,0)+(360-145:1.7)$) arc (360-145:135:1.7);

    }
  }
}

\title{Parting gravity's tail: quadrupole tails at fifth order and
  beyond via integer partitions}

\author{Alex Edison
  \orcidlink{0000-0002-5430-9500}}

\affiliation{Department of Physics
  and Astronomy, Northwestern University, Evanston, Illinois, 60208,
  USA}

\emailAdd{alexander.edison@northwestern.edu}

\abstract{This work studies the systematic organization of
  higher-order gravitational quadrupole tails using generalized
  unitarity methods imported from the study of scattering amplitudes.
  The first major result is a constructive algorithm for generic
  arbitrary-order tail effective actions which links the structure of
  their loop integral basis expansion with integer partitions, and
  predicts that only a single new unitarity cut needs to be evaluated
  at each tail order with all other contributions given in terms of
  lower-loop data.  The algorithm is employed to compute the
  tail-of-tail-of-tail-of-tail-of-tail (T$^5$) contributions to the
  effective action and associated energy loss to gravitational waves.
  Validation of the new effective action and radiated energy is done
  through counterterm extraction and renormalization analysis, leading
  to complete agreement with known counterterms and renormalization
  flow equations.
}

\begin{document}

\maketitle

\section{Introduction}

Gravitational waves and their sources hold a large and pressing
interest for both phenomenological and theoretical study.  The
LIGO-VIRGO-KAGRA network
\cite{LIGOScientific:2014pky,VIRGO:2014yos,KAGRA:2020tym} demands
more-precise predictions to inform template models, and pursuing this
theoretical precision has exposed a host of conceptual and technical
questions.

The best-explored framework to understand gravitational waves and
their binary inspiral sources is the post-Newtonian approximation of
general relativity (GR).  In this approximation, the objects under
consideration are \emph{slowly moving} and \emph{weakly
  gravitationally interacting}, which matches well with the expected
behavior of a quasi-circular inspiral phase of binary black hole
systems.  The effective theory is well-studied through both classical
perturbation theory \cite{Damour:2014jta, Blanchet:2013haa,
  Blanchet:2024mnz} and Lagrangian descriptions manifesting the
relevant degrees of freedom \cite{Goldberger:2004jt,Kol:2007rx,
  Kol:2007bc, Kol:2009mj, Kol:2010ze,Bernard:2015njp}.  Within this
effective field theory perspective, the non-linear dynamics of gravity
lead to subtle phenomena known as ``hereditary effects''
\cite{Blanchet:1992br} or ``tails'' \cite{Blanchet:1987wq}, which
account for the interaction between the residual binding-potential
modes of the gravitational field, and the radiative modes.  The tails
importantly lead to corrections to the radiated energy of a binary
black hole system, and also result in \emph{non-local-in-time}
evolution \cite{Blanchet:1992br}.  This induced history dependence of
the system complicates the connection between scattering and bound
systems at subleading perturbative orders \cite{Kaelin2020a,
  Kaelin2020, Cho:2021arx, Buonanno:2022pgc}.

In Refs.~\cite{Edison:2022cdu, Edison:2023qvg}, the current author and
Levi studied higher-order tail effects using generalized unitarity
methods.  These techniques, originally developed for the study of
scattering amplitudes in particle physics, allow for the construction
of observables using simple, gauge invariant building blocks.  They
allowed access to the tail-of-tail-of-tail-of-tail (T$^4$), the most
advanced tail ever computed for generic quadrupoles; traditional field
theory techniques have achieved T$^3$ \cite{Marchand:2016vox}, while
further orders have been computed using the ``self-force'' framework,
which is limited to an analytic expansion in the binary mass ratio
\cite{Fujita:2011zk, Fujita:2012cm, Barack:2018yvs,
  Warburton:2021kwk}.  Additionally, generalized unitarity methods
naturally organized the results in a way that highlighted recurring
patterns between tail orders.  The current work builds on the success
of Refs.~\cite{Edison:2022cdu, Edison:2023qvg} by identifying a
combinatoric organization of the generalized unitarity decomposition
of all currently known tails.  This systematic solution reduces the
difficulty of constructing new tail orders to evaluating a single new
unitarity cut diagram at each loop order; all of the rest of the
needed information is recycled from previous orders.

The structure of the paper is as follows.  \Cref{sec:rev} briefly
reviews the setup of the problem: the composite-particle EFT, in-in
formalism for dissipative actions, and unitarity-based methods.  Next,
\cref{sec:alg} introduces a computation method for arbitrary-order
tails.  The method associates the iterative structure of the effective
action to permutations of integer partitions, and exactly reproduces
all of the results of \cite{Edison:2023qvg}.  This novel method is
deployed in \cref{sec:t5} to streamline the computation of the
tail-of-tail-of-tail-of-tail-of-tail (T$^5$) effective action,
\cref{eq:t5-action}.  \Cref{sec:energy} covers the extraction of
inclusive energy loss, incorporating the new correction from T$^5$,
\cref{eq:t5-loss}, as well as verifying the renormalization group flow
of the quadrupole coupling.  A short summary of the integrals needed
for T$^5$ is given in \cref{sec:integrals}.

\section{Review}
\label{sec:rev}
Many aspects of the current work build directly on
Refs.~\cite{Edison:2023qvg, Edison:2022cdu}.  As such, we will limit
the current review to a brief recap of the necessary terminology and
symbols and direct interested readers to \prevpaper{} for a more
pedagogical discussion and contextual references.

We start from the post-Newtonian (PN) effective field theory of
general relativity
\cite{Goldberger:2004jt,Levi:2018nxp,Porto:2016pyg}, in which the
objects under consideration are \emph{slow moving} and \emph{weakly
  interacting}.  Under these assumptions, we have two perturbative
quantities -- the (relative) velocity of the objects and their
potential, governed by Newton's Constant -- which for quasi-circular
orbits can be related via the virial theorem
\begin{equation}
  v^2 \sim \frac{G_N m}{r} \ll 1\,,
\end{equation}
and can approximately separate the graviton into near zone modes
(which account for binding potentials between objects), and far zone
(radiation) modes.  However, the nonlinear nature of gravity leads to
non-trivial interference phenomena that occur at subleading orders in
perturbation theory.  The lowest-order such effect that contributes
corrections to gravitational radiation is the \emph{quadrupole tail},
in which gravitational radiation sourced from a quadrupole interacts
one (or more) times with a background gravitational potential.

From an effective field theory perspective, we study these effects by
considering an effective theory for post-Newtonian gravity coupled to
a \emph{composite particle} which admits a spatial multipole expansion
\cite{Goldberger:2009qd, Levi:2010zu, Ross:2012fc,
  Levi:2015msa,Levi:2018nxp}
\begin{align}
  S_{\text{eff}}[g_{\mu\nu},y^\mu,e_{A}^{\,\,\mu}]
  &= 
    -\frac{1}{16\pi G}\int d^4x
    \sqrt{-g}\,R\left[g_{\mu\nu}\right] +\, S_{\text{matter}}
    [g_{\mu\nu}(y),y^\mu,e_{A}^{\,\,\mu}] \\
    S_{\text{matter}}[g_{\mu\nu}(y),y^\mu,e_{A}^{\,\,\mu}]
  &= -\int dt \sqrt{g_{00}}\, \biggr[ E(t) - I^{ij}(t) \mathcal{E}_{ij} + \text{others} \biggr] \,.
\end{align}
where $E(t)$ is the gravitating energy of the system,
$\mathcal{E}_{ij}$ is the even-parity projection of the curvature
tensor, $I^{ij}(t)$ a time-dependent charge-type quadrupole, and
``others'' refers to related sources that lead to interference effects
like the ``failed tail'' \cite{Henry:2023sdy}, higher-order-multipole
tails \cite{Almeida:2021jyt}, and current-type tails, which we will
not explore here. $y^\mu$ and $e_{A}^\mu$ are the matter
(binary-as-point-particle) coordinate and associated tetrad,
respectively; they are necessary for the formal definition of the EFT,
but will play no direct role in the following computations.  From this
starting theory, we integrate out the interaction of the quadrupole
with gravity, which will produce an effective action describing the
potentially \emph{non-local-in-time} evolution of the quadrupole (in
frequency space)
\cite{Galley:2009px,Damour:2014jta,Foffa:2011np,Galley:2015kus}:
\begin{align}
 S_{\text{eff}}^{\text{tails}}
  &= \int \diff \omega\ f(\omega) \underbrace{I^{ij}(\omega)I_{ij}(-\omega)}_{\kappa(\omega)} \,.
  \label{eq:tail-eff}
\end{align}
Because these effects contain dissipative (and thus time-asymmetric)
contributions \cite{Blanchet:1987wq, Blanchet:1992br}, it is necessary
to employ an ``in-in'' or ``closed time path'' (CTP) formalism to
recover the correct causal behavior of the system \cite{Galley:2009px,
  Galley:2012qs, Galley:2012hx, Galley:2014wla, Galley:2015kus}.  The
CTP prescription is rather simple for the tails:
\begin{itemize}
\item We associate a ``history label'' in $\{+,-\}$ to each of the
  quadrupoles taking part in the tail: $I^{ij}(\omega) \rightarrow I^{ij}_{\pm}(\omega) \Rightarrow \kappa(\omega) \rightarrow \kappa_{\pm \mp}(\omega)$;
\item All radiation-mode graviton propagators use causal
  (advanced/retarded) $\ctpeps$ prescriptions in Fourier space
  according to which history labels they are connected to:
  \begin{equation}
    G_{-+/+-}(x-x') = G_{\text{ret}/\text{adv}}(x-x') = \int \frac{\Diff{D} p}{(2 \pi)^D} 
    \frac{e^{-i p_{\mu} (x-x')^{\mu}}}{(p^0 \pm  \ctpeps)^2 - |\vec{p}|^2}\,.
  \end{equation}
  Note that we will employ dimensional regularization as our
  regularization scheme, where $D$ will denote the analytically
  continued number of spacetime dimensions, and $d=D-1$ the spatial
  dimensions;  
\item Momentum conservation implies there is only one frequency in the
  tails problem. Thus, for each diagram we sum over taking all
  radiation propagators to be retarded, then all advanced;
\item The sum over propagator types leads to a modified $f(\omega)$
  from \cref{eq:tail-eff} with both $\omega$-even $\to$
  \emph{conservative} terms and $\omega$-odd $\to$ \emph{dissipative}
  terms
  \begin{equation}
    S_{\text{CTP}}^{\text{tails}}
    = \frac{1}{2} \int \diff \omega\ \left[ f_{\text{even}}(\omega) (\kappa_{-+}(\omega) + \kappa_{+-}(\omega)) + f_{\text{odd}}(\omega)(\kappa_{-+}(\omega) - \kappa_{+-}(\omega))\right]\,.
      \label{eq:ctp-tails}
  \end{equation}
\end{itemize}
The CTP prescription has associated extensions to
Noether's theorem, which we will employ in \cref{sec:energy}.

We can organize the computation of the tail effective action,
\cref{eq:tail-eff}, using generalized unitarity methods.  First, we
striate the effective action according to the loop order of diagrams
that contribute (or, equivalently, the naive powers of $G_N E$), which
we refer to as $S_{T^n}$.  Each of these loop orders is then further
decomposed into a basis of Euclidean loop momentum integrals,
$\mathcal{I}_i$, with $d$- and $\omega$-dependent coefficients
\begin{align}
  S_{T^n} = \int \diff \omega
  \sum_{\substack{i\text{ in }\\\text{integral basis}}}
  c_i(d,\omega,I^{ij}(\omega))  \mathcal{I}_i \,,
\end{align}
where all of the $\ctpeps$ prescription choices contribute through the
analytic continuation of the results of the loop integrals.  The
coefficients in this basis are computable via generalized unitarity
cuts \cite{Bern:1994zx, Bern:1994cg, Britto:2004nc,
  Anastasiou:2006jv,Britto:2007tt,Ossola:2006us, Forde:2007mi,
  Carrasco:2024knk, Bern:2024vqs}
\begin{align}
\cut_{\graph{}}
  &= \sum_{\substack{\text{states along}\\\text{ edges of }\graph{}}}
  \prod_{v \in \substack{\text{vertices}\\\text{ of }\graph{}}} \mathcal{A}_{\text{tree}}(v)  \,,
  \label{eq:cut-abs}
\end{align}
where $G$ is a graph relevant to the problem.  For gravitons, the sum
over states acting on a pair of graviton polarization tensors
$\pol_k^{\mu \nu}$ with associated momentum $k$ is implemented by the
physical state projector
\begin{align}
  \sum_{\text{states}} \pol^{\mu \nu}_k \pol^{\alpha \beta }_k 
  \equiv \mathcal{P}^{\mu \nu; \alpha \beta}_k 
  &= \frac{1}{2} \left( P_k^{\mu \alpha}P_k^{\nu \beta} + P_k^{\mu \beta}P_k^{\nu \alpha} - 
    \frac{2}{D-2} P_k^{\mu \nu}P_k^{\alpha \beta} \right) \ , 
    \label{eq:phys-proj}\\
  P^{\mu \nu}_k &\equiv \eta^{\mu \nu} - \frac{k^\mu q^\nu + k^\nu
                  q^\mu}{k \cdot q } \ ,
\end{align}
in which $q^\mu$ is a null reference vector that will drop out of
\cref{eq:cut-abs}.  In the case of tails, the amplitudes we care about
are:
\begin{itemize}
\item The quadrupole source amplitude for producing radiation gravitons,
  \begin{align}
    \mathcal{M}_{Ig}
    &= \lambda_I I^{ij}(\omega)
  (\omega_g k_g^i \pol_g^0 \pol_g^j
  + \omega_g k_g^j \pol_g^0 \pol_g^i - k_g^i k_g^j \pol_g^0 \pol_g^0 
      - \omega_g^2 \pol_g^i \pol_g^j) \notag \\
    & \equiv \lambda_I J_I^{\mu \nu} \pol_{\mu \nu} \notag \\
    &= \begin{tikzpicture}
      \node[quad src={3}{\(I^{ij}\)},outer sep=0] (a) at (0,0){};
      \node (e) at (1,0) {\(\pol_{\mu \nu}\)};
      \draw[photon big] (a)--(e);
    \end{tikzpicture} \ ,
      \label{eq:quad-src}
  \end{align}
  with coupling constant $\lambda_I$, $k_g$ and $\pol_g$ the momentum
  and polarization of the emitted graviton, and $p_I$ the momentum of
  the quadrupole which we take to be the largest scale in the problem,
  and will evaluate in its rest frame \emph{after} unitarity state
  sewing.
\item The single potential-mode source amplitude (analogous to a
  graviton coupled to a worldline \cite{Mogull2021})
  \begin{equation}
    \mathcal{M}_{E g} = \mathcal{M}_{sgs} = \frac{\lambda_E}{m_I^2} p_I^\mu p_I^\nu \pol_{\mu \nu}= \begin{tikzpicture}
      \node[mass src={2}] (a) at (0,0){};
      \node (e) at (1,0) {\(\pol_{\mu \nu}\)};
      \draw[photon big] (a)--(e);
    \end{tikzpicture} \,,
    \label{eq:pot-src}
  \end{equation}
  with coupling constant $\lambda_E$ and $m_I = p_I^0$.  While we draw
  these with the same squiggly line as the gravitons in
  \cref{eq:quad-src}, we will (after state sewing) take a
  $m_I \to \infty$ and $\omega_{\text{pot}}\to 0$ limit , projecting
  these gravitons into the $\pol^{00}$ potential mode.
\item The source contact amplitude
  \begin{align}
    \mathcal{M}_{Eg^2} = \lim_{m_I \to \infty} \msggs =
    & \frac{\lambda_g\lambda_E }{\omega_{1}^2} \frac{\delta(\omega_1-\omega_2)}{2 (k_1\cdot k_{2})} 
      \Big[ (k_1\cdot k_{2}) \pol_1^0 \pol_2^0
      + \omega_{1}( (\pol_2\cdot k_{1})\pol_1^0 \notag\\
    &- (\pol_1\cdot k_{2})\pol_2^0) - \omega_{1}^2(\pol_1\cdot \pol_{2})\Big]^2 + \mathcal{O}(m_I^{-1}) \, \notag \\
    &= \begin{tikzpicture}
      \node[mass src={2}] (a) at (0,0){};
      \node (e1) at (1,0.6) {\(\pol_1^{\mu \nu}\)};
      \node (e2) at (1,-0.6) {\(\pol_2^{\rho \sigma}\)};
      \draw[photon big] (a)--(e1);
      \draw[photon big] (a)--(e2);
    \end{tikzpicture} \ ,
      \label{eq:pot-cont}
  \end{align}
  with bulk graviton coupling constant $\lambda_g$, and kinematic data
  for two gravitons labeled 1 and 2.  The $m_I \to \infty$ limit is
  taken to remove the quantum off-shell propagation of the classical
  ``source'', and should in practice be taken after the unitarity
  sewing.  We show the result of the limit on the amplitude itself to
  highlight the expected leading-order behavior (and the double-copy
  structure \cite{Brandhuber:2021kpo}).  The drawn gravitons are both
  radiation-mode, but the amplitude itself contains a potential-mode
  pole.
\item Bulk graviton amplitudes
  \begin{equation}
    \mathcal{M}_{g^n} =
    \begin{tikzpicture}
      \node (1) at (225:1.25) {1};
      \node (n) at (315:1.25) {n};
      \node (c) at (90:1.25) {};
      \draw[photon big] (1) -- (0,0);
      \draw[photon big] (n) -- (0,0);
      \draw[photon big] (c) -- (0,0);
      \node[bulk] (b) at (0,0) {\(\mathcal{M}_{n}\)};
      \draw[dotted,thick] (215:0.75) arc (215:100:0.75);
      \draw[dotted,thick] (80:0.75) arc (80:-35:0.75);
    \end{tikzpicture} \,.
    \label{eq:bulk}
  \end{equation}
\end{itemize}
All of these amplitudes are relatively straightforward to compute
using the \inctr{}~\cite{Edison:2020ehu} Mathematica package.  The
package produces arbitrarily-normalized amplitudes, prompting the
introduction of the coupling constants in the above expressions.  We
fix the normalization conventions by matching individual pieces and
lowest-order results against
Refs.~\cite{Goldberger:2009qd,Galley:2015kus}
\begin{equation}
  \lambda_I = \sqrt{2 \pi G_N} \qquad 
  \lambda_E = - E \sqrt{8 \pi G_N} \qquad
  \lambda_g = - \sqrt{32 \pi G_N} \,,
\end{equation}
with Newton's constant $G_N$, and the total gravitating mass-energy of
the system $E$.  As constructed via \cref{eq:cut-abs}, the generalized
unitarity cuts are rational functions of loop momentum, potentially
contracted into the ``external'' quadrupoles.  In order to extract the
integral basis coefficients from them we must perform integral and
tensor reduction, keeping only terms from the integral reduction where
none of the cut propagators are collapsed.  The \emph{reduced}
generalized unitarity cut is now a linear combination of basis
integrals and coefficients; all that is left for explicit basis
coefficient extraction is to normalize the reduced cut by a symmetry
factor
\begin{align}
\cut_{\graph{}}
  \xrightarrow[\text{tens. red.}]{\text{IBPs}}
  \rcut{\graph{}} \mathcal{I}_G\,, \qquad  
  \frac{\rcut{\graph{}}}{|\graph{}|}\mathcal{I}_G
  &=  \sum_{\substack{\mathcal{I}_i \text{ has propagators}\\\text{compatible with }\graph{}}} c_i \mathcal{I}_{i} \,,
  \label{eq:cut-matching}
\end{align}
where $|\graph{}|$ is the size of the automorphism group of the graph.

We will retain the tail symmetry convention from \prevpaper{} where we
sum over using both the advanced and retarded propagator in the CTP
evaluation, but include a factor of 2 as a reflection symmetry factor
\emph{even for graphs that naively do not have this symmetry}.

\section{Conjectured iterative tail construction}
\label{sec:alg}

Ref.~\cite{Edison:2023qvg} demonstrated how generalized unitarity
methods could be efficiently applied to the case of tail effective
actions.  In that work, the minimal loop integral basis was identified
via careful analysis of possible diagram topologies. Then, the
corresponding basis coefficients were computed using generalized
unitarity cuts to sew together appropriate gravity amplitudes.  It was
observed that many of the coefficients present in T$^4$ factorized
into lower-loop-order structures, as did the integral basis itself.

In order to manifest these observed factorization patterns, we propose
the following algorithm for constructing arbitrary tail orders (T$^n$)
organized as they would be according to generalized unitarity methods.
First, we construct the basis of cut diagrams that contribute:
\begin{enumerate}
\item Enumerate the integer partitions of $n$ into $1,\dots,n$:
  \begin{equation}
    \mathfrak{p}_n = \left\{\{\partel{1}, \dots\} \Big| \sum_i \partel{i} = n
      \text{ and } x_i \in \mathbb{Z}
      \text{ and } 1 \le \partel{i} \le n
      \text{ and } \partel{i} > \partel{i+1} \right\} 
  \end{equation}
\item Find all nonequivalent permutations of the partitions, which we
  will call $\mathfrak{s}_n$
\item For each $\sigma \in \mathfrak{s}_n$ construct the cut diagram
  $G_{\sigma}$:
  \begin{enumerate}
  \item Read the elements in $\sigma$ from left to right and
    insert
  \begin{itemize}
  \item A source contact,~\cref{eq:pot-cont}, for each
    $\partel{i}=1$,;
  \item A bulk graviton contact $\mathcal{M}_{\partel{i}}$,
    \cref{eq:bulk}, for each $\partel{i}>1$, with cut propagators
    connected to $\partel{i}-2$ simple potential sources,
    \cref{eq:pot-src};
  \item A cut propagator for between each sequential element,
    \cref{eq:phys-proj};
  \end{itemize}
\item Complete the cut diagram by inserting a quadrupole
  $I^{ij}(\omega_1)$ connected by a cut propagator to the first
  element, and a $I^{mn}(\omega_2)$ connected by a cut propagator to
  the final element.  Note that momentum conservation implies
  $\omega_1 = - \omega_2$.
\end{enumerate}
\end{enumerate}
For example, the leading tail has $n=1$, for which
$\mathfrak{s}_1=\{\{1\}\}$. Step 3(a) tells us this necessitates
inserting a source contact, which is connected on either side to a
quadrupole:
\begin{equation}
  \{1\} \Rightarrow
  \begin{tikzpicture}
    \draw pic {hills={1}};
  \end{tikzpicture}\,.
  \label{eq:lead-tail}
\end{equation}
Similarly, for T$^2$ we have $\mathfrak{s}_2=\{\{2\},\{1,1\}\}$,
leading to the cut diagrams
\begin{equation}
  \{1,1\} \Rightarrow
  \begin{tikzpicture}
    \draw pic {hills={2}};
  \end{tikzpicture}\,, \qquad
  \{2\} \Rightarrow
  \begin{tikzpicture}[smooth]
      \node[quad src={3}{\(I^{ij}\)}] (a) at (0,0){};
      \node[quad src={3}{\(I^{mn}\)}] (b) at (0,3) {};
      \draw pic {mamp={a}{b}{2}}{};
    \end{tikzpicture}\,.
    \label{eq:tail2-cuts}
\end{equation}
Further examples will be provided for T$^5$ below.

The reduced unitarity cut coefficients for these basis diagrams --
normally computed via \cref{eq:cut-matching} or through reductions of
Feynman diagrams -- factorize based on the integer partition, and are
given by
\begin{equation}
  \rcut{G_{\sigma}} = \mathcal{R} \prod_{\partel{i} \in \sigma} \cutcoef{\partel{i}}
  \label{eq:tail-cut-gen}
\end{equation}
where $\mathcal{R}$ is the radiation-reaction coefficient given below,
and the $\cutcoef{i}$ will first appear in the computation of T$^i$,
but can then be recycled for higher loop orders.  From all of the cut
coefficients computed in \prevpaper, we can identify the functions
needed up through $\cutcoef{4}$ as
\begin{subequations}
  \label{eq:coef-table}
\begin{align}
  \mathcal{R} &= \omega^4 \frac{(d+1)(d-2)}{(d+2)(d-1)} \,,
                \qquad \mathcal{D}_3 = (d-3)(d-1)(d+1) \,,
                \qquad \mathcal{D}_1 = (d-1)d(d+1) \,, \\
  \cutcoef{1} &= -\frac{ \mathcal{P}_4 }{2d \mathcal{D}_3} \,, \label{eq:ccoef-1}\\
  \cutcoef{2} &= \frac{\omega^2 (2d-3) \mathcal{P}_8}{3\mathcal{D}_3 \mathcal{D}_1(d-2) (3d-2)(3d-4)} \,,\label{eq:ccoef-2}\\
  \cutcoef{3} &= \frac{\omega^2 (3d-5) \mathcal{P}_{11}}{2 \mathcal{D}_3^2 \mathcal{D}_1 (2d-3)(3d-2)(3d-4)} \,,\label{eq:ccoef-3}\\
  \cutcoef{4} &= \frac{\omega^2\mathcal{P}_{22}}{30 \mathcal{D}_3^3 \mathcal{D}_1 (d-2)^2 d (2d-3) (3d-2)^2 (3d-4)^2 (5d-6)(5d-8)} \,, \label{eq:ccoef-4}
\end{align}
\end{subequations}
with the numerator polynomials $\mathcal{P}_x$ given by
\begin{subequations}
\begin{align}
  \mathcal{P}_4
  &= 12 - 2d + 5^2 - 4d^3 + d^4 \,,\\
  \mathcal{P}_8
  & = 960 - 1696d + 424d^2 - 476d^3 + 330d^4 - 39d^5 + 53d^6 - 45d^7 + 9d^8 \,,\\
  \mathcal{P}_{11}
  &= -3024 + 3720d + 2980d^2 + 996 d^3 - 2426 d^4 - 737 d^5 + 799 d^6 - 284 d^7 - 36 d^8 \notag\\
  &\quad + 223 d^9 - 117 d^{10} + 18d^{11} \,,\\
  \mathcal{P}_{22} &= -60914073600 + 447389982720d - 1430856658944d^2 + 2688616654848d^3 \notag \\
  &\quad - 3454174264320d^4 + 3432661203456d^5- 2911608239232d^6 + 2203112130496d^7 \notag \\
  &\quad -  1503589473248d^8 + 944837229872d^9 - 555238832200d^{10} + 309327324244d^{11} \notag \\
  &\quad - 174143104426d^{12} + 101573518519d^{13} - 53485147433d^{14} + 21042703665d^{15} \notag \\
  &\quad  - 4908635433d^{16} + 14340403d^{17} + 478961469d^{18} - 198558153d^{19} + 43610967d^{20} \notag \\
  &\quad - 5364630d^{21} + 291600d^{22} \,.
\end{align}
\end{subequations}

The basis integrals themselves are in one-to-one correspondence with
the diagrams.  Unlike for the cut coefficients which factorize at each
$\partel{i}$, the integrals only properly factorize at $\partel{i}=1$
and are otherwise honest multi-loop integrals whose internal topology
is determined by $\sigma$.  We will refer to the integral factors
based on the bulk vertex structure that makes up their topology.  For
instance, the integral topology corresponding to \cref{eq:lead-tail}
is
\begin{equation}
  \begin{tikzpicture}
    \draw pic {hills={1}};
  \end{tikzpicture}
  \Rightarrow
  \left(\int \frac{\diff^d \ell}{(2 \pi)^d} \frac{1}{-\ell^2 + \omega^2}\right)^2 = \mathcal{I}_{\emptyset}^2
\end{equation}
and the one for $\{2\}$ from \cref{eq:tail2-cuts} is
\begin{align}
  \begin{tikzpicture}[smooth]
    \node[quad src={3}{\(I^{ij}\)}] (a) at (0,0){};
    \node[quad src={3}{\(I^{mn}\)}] (b) at (0,3) {};
    \draw pic {mamp={a}{b}{2}}{};
  \end{tikzpicture}
  \Rightarrow
  \int \left(\prod_{i=1}^3 \frac{\diff^d \ell_i}{(2 \pi)^d}\right)
    \frac{1}{(-\ell_1^2 + \omega^2)(-\ell_2^2 + \omega^2)(-\ell_3^2)(-(\ell_1+\ell_2+\ell_3)^2)} = \mathcal{I}_4 \,.
\end{align}
We refer to the single-propagator integral that first appears in the
radiation reaction as $\mathcal{I}_{\emptyset}$, and the next new
topology (from T$^2$) that contains a 4-point vertex as
$\mathcal{I}_4$.  Luckily, many of the multi-loop integrals are
relatively straightforward to evaluate, see \cref{sec:seven-int}, with
the first difficult integral appearing at T$^4$, see
\cref{sec:four-four-int}.

With both the coefficients and integrals, we can construct the
effective action contribution as
\begin{align}
  S_{T^n} &= \lambda_I^2 (\lambda_E \lambda_g)^n \intw
            \sum_{\sigma \in \mathfrak{s}_n} \frac{1}{|G_\sigma|}
            \left(\mathcal{R} \prod_{\partel{i} \in \sigma} \cutcoef{\partel{i}} \right)
            \mathcal{I}_{G_\sigma} \,.
            \label{eq:gen-tails}
\end{align}
Applying the above construction algorithm does, of course, exactly
reproduce the RR through T$^4$ results of \prevpaper.

We note that the primary reason that this construction algorithm
remains a conjecture is the integral basis decomposition.  It is
entirely possible that some higher-order tail will contain a necessary
basis integral that is not captured via the integer partition scheme
above.  It would be interesting to see if modern integration methods
like intersection theory \cite{Frellesvig:2020qot, Mastrolia:2018uzb}
would be able to shed light on this problem.

\section{Construction of tail-of-tail-of-tail-of-tail-of-tail (T$^5$)}
\label{sec:t5}

\begin{figure}[h]
  \begin{subfigure}[b]{0.19\textwidth}
    \centering
    \begin{tikzpicture}
      \draw pic {hills={5}};
    \end{tikzpicture}
    \caption{$\{1,1,1,1,1\}$}
    \label{fig:hills}
  \end{subfigure}
  \begin{subfigure}[b]{0.19\textwidth}
    \centering
    \begin{tikzpicture}
      \node[quad src={3}{\(I^{ij}\)}] (a) at (0,0){};
      \node[quad src={3}{\(I^{mn}\)}] (b) at (0,6) {};
      \node[mass src={2}] (m1) at (0,1){};
      \node[mass src={2}] (m2) at (0,2){};
      \node[mass src={2}] (m3) at (0,3){};
      \draw pic {mamp={m3}{b}{2}};
      \draw pic {cut hill={a}{m1}};
      \draw pic {cut hill={m1}{m2}};
      \draw pic {cut hill={m2}{m3}};
    \end{tikzpicture}
    \caption{$\{2,1,1,1\}$}
    \label{fig:four-ooo}
    
  \end{subfigure}
  \begin{subfigure}[b]{0.19\textwidth}
    \centering
    \begin{tikzpicture}
      \node[quad src={3}{\(I^{ij}\)}] (a) at (0,0){};
      \node[quad src={3}{\(I^{mn}\)}] (b) at (0,6) {};
      \node[mass src={2}] (m1) at (0,1){};
      \node[mass src={2}] (m2) at (0,2){};
      \node[mass src={2}] (m3) at (0,5){};
      \draw pic {mamp={m2}{m3}{2}};
      \draw pic {cut hill={a}{m1}};
      \draw pic {cut hill={m1}{m2}};
      \draw pic {cut hill={m3}{b}};
    \end{tikzpicture}
    \caption{$\{1,2,1,1\}$}
    \label{fig:o-four-oo}    
  \end{subfigure}
  \begin{subfigure}[b]{0.19\textwidth}
    \centering
    \begin{tikzpicture}
      \node[quad src={3}{\(I^{ij}\)}] (a) at (0,0){};
      \node[quad src={3}{\(I^{mn}\)}] (b) at (0,6) {};
      \node[mass src={2}] (m1) at (0,1){};
      \node[mass src={2}] (m2) at (0,4){};
      \node[mass src={2}] (m3) at (0,5){};
      \draw pic {mamp={m1}{m2}{2}};
      \draw pic {cut hill={a}{m1}};
      \draw pic {cut hill={m2}{m3}};
      \draw pic {cut hill={m3}{b}};
    \end{tikzpicture}
    \caption{$\{1,1,2,1\}$}
    \label{fig:oo-four-o}
  \end{subfigure}
  \begin{subfigure}[b]{0.19\textwidth}
    \centering
    \begin{tikzpicture}
      \node[quad src={3}{\(I^{ij}\)}] (a) at (0,0){};
      \node[quad src={3}{\(I^{mn}\)}] (b) at (0,6) {};
      \node[mass src={2}] (m1) at (0,3){};
      \node[mass src={2}] (m2) at (0,4){};
      \node[mass src={2}] (m3) at (0,5){};
      \draw pic {mamp={a}{m1}{2}};
      \draw pic {cut hill={m1}{m2}};
      \draw pic {cut hill={m2}{m3}};
      \draw pic {cut hill={m3}{b}};
    \end{tikzpicture}
    \caption{$\{1,1,1,2\}$}
    \label{fig:ooo-four}
  \end{subfigure}
  \caption{Diagrams for factorized contributions recycling data from
    radiation-reaction, T$^1$, and T$^2$, captioned by the permuted
    integer partition that generates them.}
  \label{fig:iter-lower}
\end{figure}
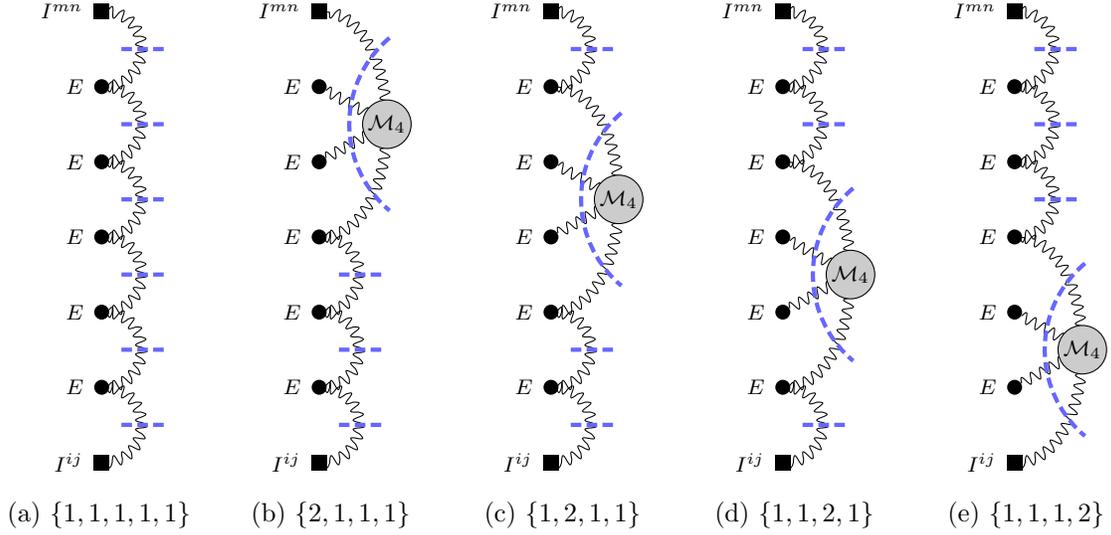

\begin{figure}[h]
  \centering
  \begin{subfigure}[b]{0.3\textwidth}
    \centering
    \begin{tikzpicture}
      \node[quad src={3}{\(I^{ij}\)}] (a) at (0,0){};
      \node[quad src={3}{\(I^{mn}\)}] (b) at (0,6) {};
      \node[mass src={2}](m1) at (0,1){};
      \node[mass src={2}](m2) at (0,2){};
      \draw pic {mamp={m2}{b}{3}};
      \draw pic {cut hill={a}{m1}};
      \draw pic {cut hill={m1}{m2}};
    \end{tikzpicture}
    \caption{$\{3,1,1\}$}
    \label{fig:five-oo}
  \end{subfigure}
  \begin{subfigure}[b]{0.3\textwidth}
    \centering
    \begin{tikzpicture}
      \node[quad src={3}{\(I^{ij}\)}] (a) at (0,0){};
      \node[quad src={3}{\(I^{mn}\)}] (b) at (0,6) {};
      \node[mass src={2}](m3) at (0,4){};
      \node[mass src={2}](m4) at (0,5){};
      \draw pic {mamp={a}{m3}{3}};
      \draw pic {cut hill={m3}{m4}};
      \draw pic {cut hill={m4}{b}};
    \end{tikzpicture}
    \caption{$\{1,1,3\}$}
    \label{fig:o-five-o}
  \end{subfigure}
  \begin{subfigure}[b]{0.3\textwidth}
    \centering
    \begin{tikzpicture}
      \node[quad src={3}{\(I^{ij}\)}] (a) at (0,0){};
      \node[quad src={3}{\(I^{mn}\)}] (b) at (0,6) {};
      \node[mass src={2}](m1) at (0,1){};
      \node[mass src={2}](m4) at (0,5){};
      \draw pic {mamp={m1}{m4}{3}};
      \draw pic {cut hill={a}{m1}};
      \draw pic {cut hill={m4}{b}};
    \end{tikzpicture}
    \caption{$\{1,3,1\}$}
    \label{fig:oo-five}
  \end{subfigure}
  \caption{Diagrams that recycle the single $\mathcal{M}_5$ bulk
    contribution first appearing at T$^3$, captioned by the permuted
    integer partition that generates them.}
  \label{fig:iter5}
\end{figure}
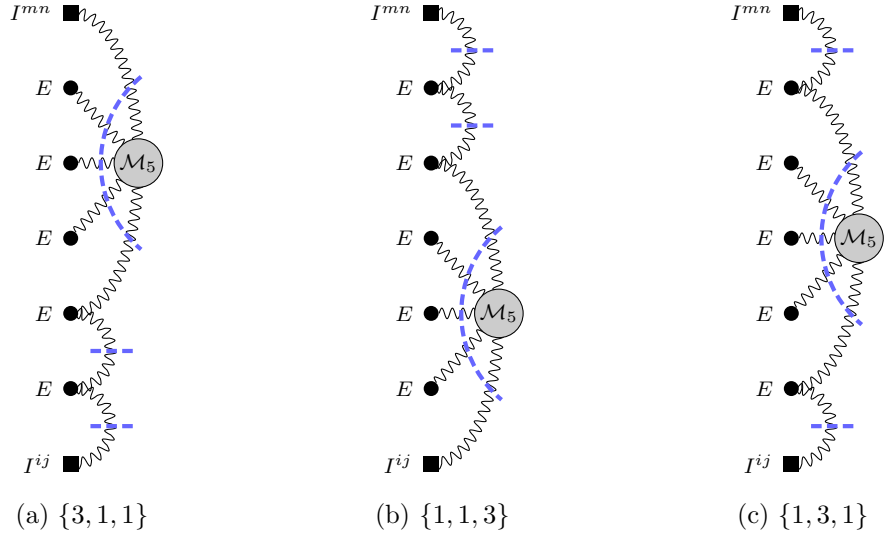

\begin{figure}
  \centering
  \pgfmathsetmacro{\mampPushback}{1.6}
  \begin{subfigure}[b]{0.2\textwidth}
    \centering
    \begin{tikzpicture}
      \node[quad src={3}{\(I^{ij}\)}] (a) at (0,0){};
      \node[quad src={3}{\(I^{mn}\)}] (b) at (0,6) {};
      \node[mass src={2}] (m1) at (0,5){};
      \draw pic {mamp={a}{m1}{4}};
      \draw pic {cut hill = {m1}{b}};
    \end{tikzpicture}
    \caption{$\{1,4\}$}
    \label{fig:o-six}
  \end{subfigure}
  \begin{subfigure}[b]{0.2\textwidth}
    \centering
    \begin{tikzpicture}
      \node[quad src={3}{\(I^{ij}\)}] (a) at (0,0){};
      \node[quad src={3}{\(I^{mn}\)}] (b) at (0,6) {};
      \node[mass src={2}] (m1) at (0,1){};
      \draw pic {mamp={m1}{b}{4}};
      \draw pic {cut hill = {a}{m1}};
    \end{tikzpicture}
    \caption{$\{4,1\}$}
    \label{fig:six-o}
  \end{subfigure}
  \pgfmathsetmacro{\mampPushback}{1.2}
  \begin{subfigure}[b]{0.2\textwidth}
    \centering
    \begin{tikzpicture}
      \node[quad src={3}{\(I^{ij}\)}] (a) at (0,0){};
      \node[quad src={3}{\(I^{mn}\)}] (b) at (0,6) {};
      \node[mass src={2}](m1) at (0,5){};
      \draw pic{split={a}{2}{2}{m1}};
      \draw pic{cut hill={m1}{b}};
    \end{tikzpicture}
    \caption{$\{1,2,2\}$}
    \label{fig:off}
  \end{subfigure}
  \begin{subfigure}[b]{0.2\textwidth}
    \centering
    \begin{tikzpicture}
      \node[quad src={3}{\(I^{ij}\)}] (a) at (0,0){};
      \node[quad src={3}{\(I^{mn}\)}] (b) at (0,6) {};
      \node[mass src={2}](m1) at (0,1){};
      \draw pic{split={m1}{2}{2}{b}};
      \draw pic{cut hill={a}{m1}};
    \end{tikzpicture}
    \caption{$\{2,2,1\}$}
    \label{fig:ffo}
  \end{subfigure}
  \caption{Diagrams whose factorization contains terms first appearing
    in $T^4$, captioned by the permuted integer partition that
    generates them.}
  \label{fig:t4}
\end{figure}
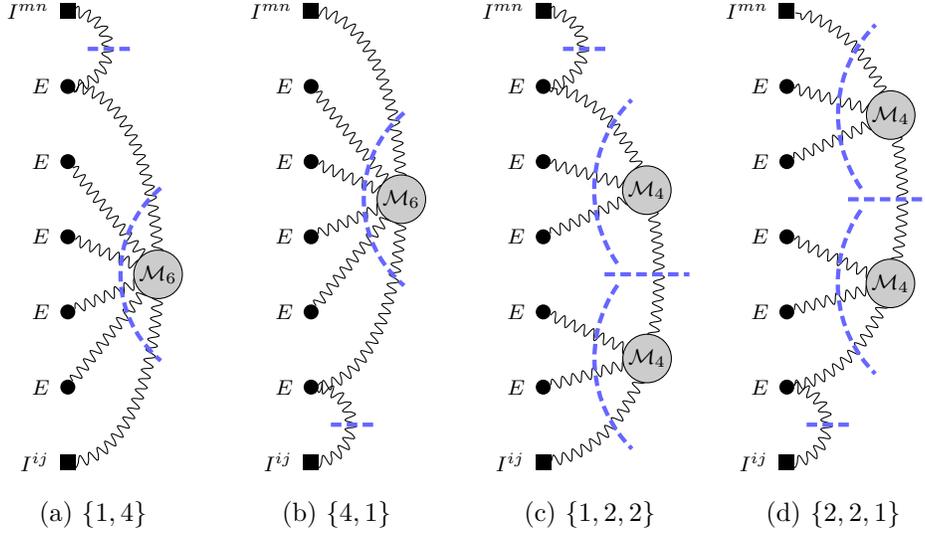

\begin{figure}[h]
  \centering
  \begin{subfigure}[b]{0.2\textwidth}
    \centering
    \begin{tikzpicture}
      \node[quad src={3}{\(I^{ij}\)}] (a) at (0,0){};
      \node[quad src={3}{\(I^{mn}\)}] (b) at (0,6) {};
      \draw pic{split={a}{3}{2}{b}};
    \end{tikzpicture}
    \caption{$\{2,3\}$}
    \label{fig:three-two}
  \end{subfigure}
  \begin{subfigure}[b]{0.2\textwidth}
    \centering
    \begin{tikzpicture}
      \node[quad src={3}{\(I^{ij}\)}] (a) at (0,0){};
      \node[quad src={3}{\(I^{mn}\)}] (b) at (0,6) {};
      \draw pic{split={a}{2}{3}{b}};
    \end{tikzpicture}
    \caption{$\{3,2\}$}
    \label{fig:two-three}
  \end{subfigure}
   \begin{subfigure}[b]{0.2\textwidth}
    \centering
    \begin{tikzpicture}
      \node[quad src={3}{\(I^{ij}\)}] (a) at (0,0){};
      \node[quad src={3}{\(I^{mn}\)}] (b) at (0,6) {};
      \node[mass src={2}](mm) at (0,3){};
      \draw pic{mamp={a}{mm}{2}};
      \draw pic{mamp={mm}{b}{2}};
    \end{tikzpicture}
    \caption{$\{2,1,2\}$}
    \label{fig:fourfour}
  \end{subfigure}
  \begin{subfigure}[b]{0.2\textwidth}
    \centering
    \pgfmathsetmacro{\mampPushback}{1.6}
    \begin{tikzpicture}
      \node[quad src={3}{\(I^{ij}\)}] (a) at (0,0){};
      \node[quad src={3}{\(I^{mn}\)}] (b) at (0,6) {};
      \draw pic{mamp={a}{b}{5}};
    \end{tikzpicture}
    \pgfmathsetmacro{\mampPushback}{1.2}
    \caption{$\{5\}$}
    \label{fig:seven}
  \end{subfigure}
  \caption{Diagrams newly appearing at $T^5$, captioned by the permuted
    integer partition that generates them.}
  \label{fig:t5-all}
\end{figure}
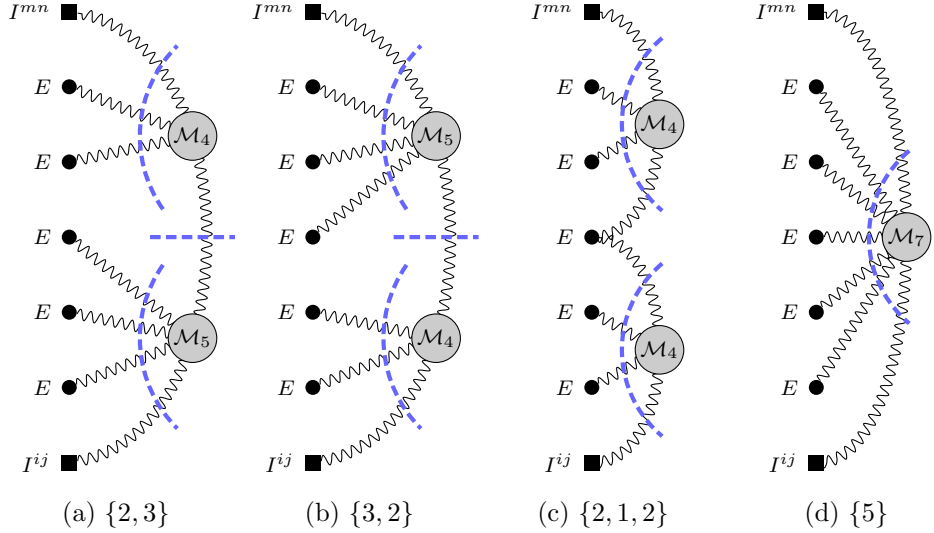

We now apply the tail construction algorithm to the case of T$^5$.
First, we identify the nonequivalent permuted partitions of 5:
\begin{equation}
  \mathfrak{s}_3 = \left\{
  \arraycolsep=12pt
  \begin{array}{llll}
    \{1,1,1,1,1\} & \{3,1,1\} & \{4,1\} & \{2,3\} \\
    \{2,1,1,1\} & \{1,3,1\} & \{1,4\} & \{3,2\} \\
    \{1,2,1,1\} & \{1,1,3\} & \{1,2,2\} & \{2,1,2\}\\
    \{1,1,2,1\} & & \{2,2,1\} & \{5\}\\
    \{1,1,1,2\} 
  \end{array}
  \right\} \,.
\end{equation}
From these partitions, we generate the actual diagrams.  The first
column is in \cref{fig:iter-lower}, the second in \cref{fig:iter5},
the third in \cref{fig:t4}, and finally the fourth in
\cref{fig:t5-all}.  The groupings are based on which tail order first
produces the most complicated sub-topology.

The first three groups are all factorizing diagrams: they rely solely on
data from previous loop orders as part of their computation.  We start
with the first group, \cref{fig:iter-lower}.  There are two distinct
contributions in this group, \cref{fig:hills} and
\cref{fig:four-ooo,fig:o-four-oo,fig:oo-four-o,fig:ooo-four}.
The unitarity cut coefficients for these diagrams are simply
\begin{align}
  \rcut{\{1,1,1,1,1\}} &= \rcut{\text{\cref{fig:hills}}}
  =  \mathcal{R}\ \cutcoefr{1}^5 \label{eq:hills}\\
  \rcut{\{2,1,1,1\}} &= \rcut{\text{\cref{fig:four-ooo}}}
  =
    \rcut{\text{\cref{fig:o-four-oo}}} =
    \rcut{\text{\cref{fig:oo-four-o}}} =
    \rcut{\text{\cref{fig:ooo-four}}} =
    \mathcal{R}\ \cutcoefr{1}^3\ \cutcoefr{2}
    \label{eq:it-four}
\end{align}
with corresponding integrals and symmetry factors
\begin{align}
  \mathcal{I}_{\text{\cref{fig:hills}}}
  &= \mathcal{I}_{\emptyset}^5 \label{eq:hills-int}\\
  |G_{\text{\cref{fig:hills}}}| &= 2 \\
  \mathcal{I}_{\text{\cref{fig:four-ooo}}}
  &=\mathcal{I}_{\text{\cref{fig:o-four-oo}}}
    =\mathcal{I}_{\text{\cref{fig:oo-four-o}}}
    = \mathcal{I}_{\text{\cref{fig:ooo-four}}}
    = \mathcal{I}_4\ \mathcal{I}_{\emptyset}^3 \label{eq:four-int}\\ 
  |G_{\text{\cref{fig:four-ooo}}}|
  &= |G_{\text{\cref{fig:o-four-oo}}}|
    = |G_{\text{\cref{fig:oo-four-o}}}|
    = |G_{\text{\cref{fig:ooo-four}}}| = 4 \,.
\end{align}
As noted above, explicit expression for these integrals in terms of
the dimensional regulator $d$ are given in \cref{sec:seven-int}.
The symmetry factors for
\cref{fig:four-ooo,fig:o-four-oo,fig:oo-four-o,fig:ooo-four} stems
from treating the potential modes and sources partaking in the bulk
interaction as indistinguishable.  The second set of diagrams,
\cref{fig:iter5}, contains only diagrams which only have a single bulk
$\mathcal{M}_5$ graviton vertex.  As such, all three have identical
unitarity cut coefficients,
\begin{equation}
  \rcut{\{3,1,1\}} = 
  \rcut{\text{\cref{fig:five-oo}}}  =
  \rcut{\text{\cref{fig:o-five-o}}}  =
  \rcut{\text{\cref{fig:oo-five}}}  =
  \mathcal{R}\ \cutcoefr{1}^2\ \cutcoefr{3} \,,
  \label{eq:five}
\end{equation}
as well as basis integrals and symmetry factors
\begin{align}
  \mathcal{I}_{\text{\cref{fig:five-oo}}}
  &=\mathcal{I}_{\text{\cref{fig:oo-five}}}
    = \mathcal{I}_{\text{\cref{fig:o-five-o}}} = \mathcal{I}_{5}\ \mathcal{I}_{\emptyset}^2 \label{eq:five-int}\\
  |G_{\text{\cref{fig:five-oo}}}|
  &= |G_{\text{\cref{fig:oo-five}}}|
    = |G_{\text{\cref{fig:o-five-o}}}| = 2 \times 3!
\end{align}

The third group, shown in \cref{fig:t4}, contains terms first seen in
T$^4$.  Their unitarity cut coefficients are easily computed as
\begin{align}
  \rcut{\text{\cref{fig:o-six}}}
  &=\rcut{\text{\cref{fig:six-o}}}
  =  \mathcal{R}\ \cutcoefr{1}\ \cutcoefr{4} \label{eq:six}\\
  \rcut{\text{\cref{fig:off}}}
  &=
    \rcut{\text{\cref{fig:ffo}}} =
    \mathcal{R}\ \cutcoefr{1}\ \cutcoefr{2}^2 \,.
    \label{eq:four}
\end{align}
The integrals and symmetry factors for \cref{fig:o-six,fig:six-o} are
also straightforward to compute
\begin{align}
  \mathcal{I}_{\text{\cref{fig:o-six}}} &= \mathcal{I}_{\text{\cref{fig:six-o}}} = \mathcal{I}_{\emptyset}\ \mathcal{I}_6\\
  |G_{\text{\cref{fig:o-six}}}| &= |G_{\text{\cref{fig:six-o}}}| = 2 \times 4! \,.
\end{align}
However, the integral-related contributions from \cref{fig:off,fig:ffo},
\begin{align}
  \mathcal{I}_{\text{\cref{fig:off}}} &= \mathcal{I}_{\text{\cref{fig:ffo}}} = \mathcal{I}_{\emptyset}\ \mathcal{I}_{4 \otimes 4}\\
  |G_{\text{\cref{fig:off}}}| &= |G_{\text{\cref{fig:ffo}}}| = 2 \times 2 \times 2 \,,
\end{align}
are more interesting because they mark the appearance of two different
features of higher-order tails.  First, these two diagrams share an
integer partition -- and thus unitarity coefficient -- with
\cref{fig:fourfour}, but differ at the level of their basis integrals.
Second, the $\mathcal{I}_{4 \otimes 4}$ integral that specifically
appears in \cref{fig:ffo,fig:off} is nontrivial to evaluate in terms
of generic functions of $d$.  We instead rely on numerical evaluation,
see \cref{sec:four-four-int}.

We arrive at the final group, \cref{fig:t5-all}.  These are
the diagrams which are in various senses ``new'' at T$^5$.
\Cref{fig:three-two,fig:two-three} follow naturally from the way that
$\{2,2\}$ worked at T$^4$.  Their unitarity coefficients are given by
\begin{equation}
  \rcut{\text{\cref{fig:three-two}}} = \rcut{\text{\cref{fig:two-three}}}
  = \mathcal{R}\ \cutcoefr{2}\ \cutcoefr{3}
\label{eq:two-three-coef}
\end{equation}
with integrals and symmetries
\begin{align}
  \mathcal{I}_{\text{\cref{fig:three-two}}}
  &= \mathcal{I}_{\text{\cref{fig:two-three}}} = \mathcal{I}_{4 \otimes 5}\\
  |G_{\text{\cref{fig:three-two}}}|
  &= |G_{\text{\cref{fig:two-three}}}| = 2 \times 2 \times 3! \,.
\end{align}
The novelty for these terms is the $\mathcal{I}_{4 \otimes 5}$ which
we again turn to numerical methods to evaluate, see
\cref{sec:four-five-int}.  Next, \cref{fig:seven}, whose interest we
alluded to above.  This diagram is the first case of two non-trivial
topologies joining at a source contact, and thus the first instance of
the order of the permuted integer partition actually mattering.
As \cref{fig:off,fig:ffo} above, the unitarity
coefficient is given by
\begin{equation}
  \rcut{\text{\cref{fig:fourfour}}} =
  \mathcal{R}\ \cutcoefr{1}\ \cutcoefr{2}^2
\end{equation}
but the integral for this diagram actually factorizes,
\begin{align}
  \mathcal{I}_{\text{\cref{fig:fourfour}}} &= \mathcal{I}_4^2 \\
  |G_{\text{\cref{fig:fourfour}}}| &= 2 \times 2 \times 2 \,,
\end{align}
unlike the other two.  In fact, understanding this distinction between
\cref{fig:fourfour} and \cref{fig:off,fig:ffo} was key to arriving at
\cref{eq:tail-cut-gen,eq:gen-tails}.

Last, but certainly not least, we arrive at the one diagram that
cannot be computed from lower-loop data, \cref{fig:seven}.  The basis
integral and symmetry factor for this diagram are simple enough,
\begin{align}
  \mathcal{I}_{\text{\cref{fig:seven}}} &= \mathcal{I}_7\\
  |G_{\text{\cref{fig:seven}}}| &= 2 \times 5! \,,
\end{align}
but the unitarity coefficient must be newly computed.  To do so, we need to compute
\begin{equation}
  \cut_{\text{\cref{fig:seven}}} = \lambda_I^2 J^{\mu \nu} J^{\rho \sigma} \left(\prod_{i=1}^{5} \mathcal{M}_{Eg}^{\alpha_i \beta_i} P^{\alpha_i \beta_i;\gamma_i \delta_i}\right) P^{\mu \nu;\gamma_0 \delta_0} P^{\rho \sigma;\gamma_6 \delta_6 }
  \mathcal{M}_{g^7}^{\gamma_0 \delta_0 \dots \gamma_6 \delta_6}
\end{equation}
using \cref{eq:phys-proj,eq:quad-src,eq:pot-src,eq:bulk}.  The
numerator for the tree amplitude has hundreds of thousands of terms,
which is further dressed with 7200 distinct combinations of
relabelings and propagators.  As such, it was vitally important to
employ both the tensor reduction method of \prevpaper{} and the state
sewing method of Ref.~\cite{Kosmopoulos:2020pcd} to reduce the
proliferation of extraneous terms.  After constructing the
loop-momentum-dependent unitarity cut, we need to reduce it to the
integral basis using integration by parts relations.  We employ a
lightly-modified version of FIRE 6.5 \cite{Smirnov:2023yhb} to enact
the integral reduction, and then extract the coefficient of the
$\mathcal{I}_7$ basis integral.  This step in the process also
provides a cross-check of the proposed integral basis: in general we
expect the reduction of this diagram \emph{without cut conditions} to
produce terms for every element of the integral basis, providing an
opportunity to identify any missing integrals.  However, the only term
which will be correct is the one whose final propagator structure
matches the cut topology, $\cutcoef{5}\ \mathcal{I}_7$.  Additionally,
the \cref{fig:three-two,fig:two-three} topologies are factorization
channels of \cref{fig:seven}, and thus the reduction of
$\cut_{\text{\cref{fig:seven}}}$ will actually produce the correct
coefficients for them as well, up to relative symmetry factors.  At
the conclusion of this process, we find the cut coefficient
\begin{align}
   \rcut{\text{\cref{fig:seven}}} = \frac{\mathcal{R}\ \mathcal{P}_{25}}
  {6 \mathcal{D}_3^4 \mathcal{D}_1 (d-2) d (2d-3)(3d-5)(5d-6)(5d-8)(3d-2)^2(3d-4)^2}
  \label{eq:seven-coef}
\end{align}
with numerator polynomial
\begin{align}
  \mathcal{P}_{25} &= 570621542400 - 3817851770880 d + 10461421320192 d^{2} \notag \\
                   &\quad - 15448588835328 d^{3} + 14470772087808 d^{4} - 11533634757504 d^{5} \notag \\
                   &\quad + 10359103552000 d^{6} - 8174350894336 d^{7} + 3835795527664 d^{8} \notag \\
                   &\quad - 412015363192 d^{9} - 640904151652 d^{10} + 441788353428 d^{11} \notag \\
                   &\quad - 49414187298 d^{12} - 213527363567 d^{13} + 340041180553 d^{14} \notag \\
                   &\quad - 370751638979 d^{15} + 309491168905 d^{16} - 190018152600 d^{17} \notag \\
                   &\quad + 83761403414 d^{18} - 26470493490 d^{19} + 6063243616 d^{20} \notag \\
                   &\quad - 1017098367 d^{21} + 119676213 d^{22} - 7235055 d^{23} - 289575 d^{24} \notag \\
                   &\quad +  60750 d^{25} \,.
\end{align}

Finally, we can combine all of the contributions using
\cref{eq:gen-tails}, evaluate all of the integrals using
\cref{sec:integrals}, and series expand in the dimension regularization
parameter $d \to 3 - 2 \dimreg$ to arrive at\footnote{Note that this
  is a different dimensional regularization scheme choice than
  \prevpaper, which uses $d \to 3 + \dimreg$.  Such differences will
  drop out of the final observables.}
\begin{align}
  S_{\text{T}^5}
  &= -\frac{1}{5} \frac{107^2}{105^2} G_N^6 E^5 \intw \omega^{10} \kappa_{-+}(\omega) \Bigg\{
    \frac{1}{6 \dimreg^3}
    + \frac{1}{\dimreg^2} \left[ \frac{3709423}{686940}-\lpegm + i \pi \sgn(\omega) \right] \notag \\
  &\quad + \frac{1}{\dimreg}\bigg[
    \frac{305250541109}{3029405400} - \frac{49}{2} \zeta_2 - \frac{1680}{107} \zeta_3 + 3 \lpegm^2 - \frac{3709423}{114490} \lpegm  \notag \\
    &\qquad + i \pi \sgn(\omega)\left(\frac{3709423}{114490} - 6\lpegm \right)
    \bigg] \notag \\
  &\quad + \bigg[ \text{\emph{Real} and $\omega$-\emph{even} terms dependent on } \mathcal{I}_{4 \otimes 4}\Big|_{\dimreg^3}\notag \\
  &\qquad + i \pi \sgn(\omega)\bigg(
    \frac{305250541109}{504900900} - 75 \zeta_2 - \frac{10080}{107} \zeta_3
    - \frac{11128269}{57245} \lpegm \notag \\
  &\quad \qquad+ 18 \lpegm^2    \bigg)
    \bigg] \notag \\
    &\quad + \odimreg{} \Bigg\} \,,
  \label{eq:t5-action}
\end{align}
where $\zeta_i$ is the Riemann zeta values, $\gamma_E$ is Euler's gamma
constant, and $\mu$ is the renormalization scale associated to
dimensional regularization.  Notably, every single individual cut's
contribution to the effective action begins at \odimreg{5}.  It is
highly nontrivial, and a mark of internal consistency, that both the
leading and subleading divergences cancel between all of the
contributions.

\section{Radiated energy and renormalization of the quadrupole coupling}
\label{sec:energy}

We employ the CTP energy extraction method, explained in
Refs.~\cite{Galley:2012hx, Galley:2014wla,Edison:2023qvg,
  Galley:2009px, Galley:2012qs}, to the T$^5$ CTP effective action,
\cref{eq:t5-action}.  Within the framework of quadrupole tails, the
process amounts to extracting the $\omega$-\emph{odd} part of the
effective action integrand, \cref{eq:ctp-tails}, and then multiplying
the resulting term by $(-i \omega)$
\begin{align}
  \Delta E
  & =
    \int \diff \omega \Big [(- i \omega)f_{\text{odd}}(\omega) \underbrace{I^{ij}(\omega)I_{ij}(-\omega)}_{\kappa(\omega)} \Big]\,.
\end{align}
In the current case of T$^5$ (as well as all previous odd-order tails
\cite{Edison:2023qvg, Edison:2022cdu,Galley:2015kus}), the CTP energy
extraction is exactly equivalent to simply taking the coefficient of
$i \sgn(\omega)$ in the integrand, multiplying it by $\omega$, and
then sending $\kappa_{-+}(\omega) \to 2 \kappa(\omega)$ (where
$\kappa(\omega)$ is the contraction of the physical quadrupoles).
Following through this process, we arrive at
\begin{align}
  \deltaE{}{T$^5$} = - \frac{1}{5} \frac{107^2}{105^2} &G_N^6 E^5
  \inte{11}
  \Bigg\{
    \frac{1}{\dimreg^2}
  + \frac{1}{\dimreg}\bigg[\frac{3709423}{114490} - 6\lpegm\bigg] \notag \\
  &+ \bigg[
  \frac{305250541109}{504900900}
  - 75 \zeta_2
  - \frac{10080}{107} \zeta_3 \notag \\
 &\quad - \frac{11128269}{57245} \lpegm
  + 18 \lpegm^2  
  \bigg] + \odimreg{1}
    \Bigg\} \,.
\label{eq:rg-t5}
\end{align}
We next need to construct the $\mathcal{O}(G_N^6)$-inclusive energy
dissipation and remove the dimensional regularization poles by
renormalizing the quadrupole coupling.

The first step of this process is to lift the quadrupole coupling to a
function of $\dimreg$ and the other parameters of the problem.  In a
slight abuse of notation, we will attach this coupling modification to
$\kappa(\omega)$ rather than $\lambda_I$:
\begin{equation}
  \kappa(\omega) \to \kappa'(\omega) \equiv
  \kappa(\omega,\mu) Z(d,\omega,G_N,E)\,.
\end{equation}
$Z(d,\omega,G_N,E)$ is a polynomial in $\omega$, $G_N$, $E$, and
$(d-3)^{-1}$ whose coefficients are fixed by demanding that the
physical observable is free of dimensional regulator divergences.
Note that we have also introduced a $\mu$-dependence to $\kappa$,
which will be relevant for studying the renormalization group flow
below.  Ref.~\cite{Edison:2023qvg} identified $Z(d,\omega,G_N,E)$
through $\mathcal{O}(G_N^4)$ as\footnote{Note that
  Ref.~\cite{Edison:2023qvg} used the dimensional regularization
  scheme $d \to 3 + \dimreg$, which we have lifted to a
  scheme-agnostic statement here.}
\begin{align}
  \kappa'(\omega) \equiv\kappa(\omega,\mu) \bigg(&
    1 + 2\frac{107}{105} \frac{(G_N E \omega)^2}{(d-3)}
    \left(1 + \frac{107}{105}\frac{(G_N E \omega)^2}{(d-3)}\right) \notag \\
    &+ \frac{1695233}{105^3} \frac{(G_N E \omega)^4}{(d-3)}
    + \mathcal{O}(G_N^5)
      \bigg)
      \,.
      \label{eq:ren-kappa}
\end{align}
In principle, there could be a new contribution at
$\mathcal{O}(G_N^5)$ which would only be detectable via the new T$^5$
terms.  Assembling the total energy dissipation of the tails,
including the leading order radiation-reaction term $\deltaE{}{RR}$,
via
\begin{equation}
  \delEInc{T$^5$} = \left[\deltaE{}{RR} + \deltaE{}{T}
    + \deltaE{}{T$^2$}+\deltaE{}{T$^3$}+\deltaE{}{T$^4$}+
    \deltaE{}{T$^5$}\right]\bigg|_{\kappa \to \kappa'}
\end{equation}
we find that all of the divergences cancel through
$\mathcal{O}(G_N^6)$ without needing to introduce a
$\mathcal{O}(G_N^5)$ correction. The leading \odimreg{-2} divergence
cancels against \deltaE{}{T} augmented by the $\frac{G_N^4}{(d-3)^2}$
from \cref{eq:ren-kappa}.  The cancellation of the subleading
divergence results from a more-complicated interplay between
\deltaE{}{T} and \deltaE{}{T$^3$} and the remaining terms of
\cref{eq:ren-kappa}.  Having dealt with the divergences, we find the
physical energy loss through $G_N^6$ to be
\begin{align}
  \delEInc{T$^5$} = \delEInc{T$^4$} + G_N^6 E^5\frac{2}{5} \frac{107^2}{105^2} &\intem{11}
  \Bigg[
  -\frac{1379886245}{10098018}
  + \frac{1680}{107} \zeta_3
  + 12 \zeta_2 \notag \\
  &- 2 \lpegm^2
  + \frac{24905541}{801430} \lpegm
  \Bigg]
  \label{eq:t5-loss}
\end{align}
where \delEInc{T$^4$} was derived in Ref.~\cite{Edison:2023qvg}.

Finally, we look at the RG flow of the now $\mu$-dependent $\kappa$.
As an observable, the energy loss must be independent of the reference
energy scale $\mu$.  Thus, the T$^5$ energy loss must obey
\begin{equation}
  \frac{\diff}{\diff \mu} \delEInc{T$^5$} = 0 + \mathcal{O}(G_N^7) \,,
  \label{eq:rg-abs}
\end{equation}
generating a classical renormalization group flow.  Since there was no
new contribution to the renormalized source coupling due to T$^5$, we
should also suspect there to be no $\mathcal{O}(G_N^5)$ contribution
to the RG flow of $\kappa(\omega,\mu)$.  Indeed, we find that
\cref{eq:rg-abs} leads to an RG equation for $\kappa(\omega,\mu)$
\begin{equation}
  \frac{\diff}{\diff \log \mu} \kappa(\omega,\mu) =
  -(2 G_N E \omega)^2 \kappa(\omega,\mu) \left(\frac{107}{105} + \frac{1695233}{105^3} (G_N E \omega)^2 \right) + \mathcal{O}(G_N^6)
  \label{eq:rg-flow}
\end{equation}
in exact agreement with Ref.~\cite{Edison:2023qvg}, but additionally
establishing the lack of $G_N^5$ corrections.

\section{Conclusion}
\label{sec:conclusion}
This work proposed a tail construction algorithm based on permuted
integer partitions, \cref{eq:gen-tails}, and applied it to the
computation of T$^5$.  This necessitated computing a novel generalized
unitarity basis coefficient, \cref{eq:seven-coef}.  The new effective
action contribution, \cref{eq:t5-action}, led to a corresponding
correction to the gravitational wave energy flux, \cref{eq:t5-loss}.

The permuted integer partition algorithm developed here exploits the
organization of the problem using generalized unitarity methods to
systematically recycle lower-loop data and identify the minimal
\emph{new} data needed at each loop order.  Should the algorithm
continue to hold at higher loops, it predicts that there is exactly
one new integral coefficient needed at each loop order.  The novel
6-loop coefficient was computed via sewing a seven-point bulk graviton
amplitude to the relevant source-analogous amplitudes,
as seen in \cref{fig:seven}.

Combining the new coefficient with the iteration terms produced the
6-loop contribution to the tail effective action.  This effective
action encodes, among other dynamics, the energy loss of the
quadrupole into the gravitational field, which by energy balance is
opposite to the energy radiated in the form of gravitational waves.
Notably, the new contribution coming from T$^5$, \cref{eq:t5-loss}, is
\emph{simpler} (in terms of $\zeta$ values which appear) than the
T$^4$ contribution found in Ref.~\cite{Edison:2023qvg} -- a similar
but less-stark difference exists between T$^2$ and T$^3$.  The new
radiated energy contribution leads to a renormalization group flow of
the quadrupole coupling, \cref{eq:rg-t5}, in exact agreement with the
RG equation found in Ref.~\cite{Edison:2023qvg}.

From a theoretical standpoint, it would be interesting to proceed to
the next order, T$^6$, both to verify the continued validity of the
proposed construction algorithm, and also because there will almost
certainly be new contributions to the renormalization and RG flow of
the quadrupole coupling, \cref{eq:rg-flow,eq:ren-kappa}.  However, it
seems difficult that this step could be achieved simply by repeating
the cut sewing and reduction used here, as it suffers from
worse-than-factorial growth in many stages of the computation.  A
particularly useful bypass of this problem would be understanding the
relationships between the $\cutcoef{i}$ in order to find a method of
constructing them directly.  A tantalizing hint in this direction is
that the irreducible polynomials appearing in the numerators of
\cref{eq:ccoef-1,eq:ccoef-2,eq:ccoef-3,eq:ccoef-4,eq:seven-coef}
double in degree when going from odd to even order, but only increase
their degree by 3 when going from even to odd.

There are also a number of interesting followup directions that do not
necessitate knowing \cutcoef{6}.  For instance, \cref{eq:gen-tails}
provides a route to resumming various sub-classes of terms whose
constituent \cutcoef{i} and $\mathcal{I}_{G}$ are already known.
Analyzing such terms and comparing them with other resummation
schemes, \eg eikonal, could prove enlightening.  Additionally, it
would be worthwhile to examine how the more subtle memory effects
\cite{Porto:2024cwd,Trestini2023, Trestini2023a, Leandro:2020txb,
  Strominger:2014pwa, Wiseman:1991ss, Thorne:1992sdb, Favata:2008yd,
  Favata:2011qi} can be captured using these types of generalized
unitarity methods.  Finally, since this framework for computing tails
only relies on the validity of the multipole expansion of the
stress-energy tensor -- with no explicit reference to a binary system
-- it would be worthwhile to explore the connection with other methods
for studying compact gravitating objects, like black hole perturbation
theory or deformations of neutron stars.

\begin{acknowledgments}
  I would like to thank Mich\`{e}le Levi, Julio Parra-Martinez, Nic Pavao,
  Fei Teng, Radu Roiban, and John Joseph Carrasco for discussion on
  this and related projects.
  I am grateful to Sasank Chava, Julio Parra-Martinez, Nic Pavao, and
  Radu Roiban for feedback on the manuscript.
  I would also like to thank Alexander Smirnov for providing guidance
  on modifying FIRE to accommodate this work's integral reduction problem.

  This research was supported by the US DOE under
  contract DE-SC0015910 and by Northwestern University via the
  Amplitudes and Insight Group, Department of Physics and Astronomy,
  and Weinberg College of Arts and Sciences.
  This research would not have been possible without the
  computational resources and staff contributions provided for the
  Quest high performance computing facility at Northwestern University
  which is jointly supported by the Office of the Provost, the Office
  for Research, and Northwestern University Information Technology.

  fill\TeX was used as part of writing the bibliography
  \cite{2017JOSS....2..222G}.
\end{acknowledgments}

\appendix
\section{Integrals}
\label{sec:integrals}
Due to the factorization of the basis integrals, all but two of the
needed integral evaluations ($\mathcal{I}_{4\otimes 5}$ and
$\mathcal{I}_{7}$) can be recycled from lower loop orders.
$\mathcal{I}_{4\otimes 5}$ requires numerical evaluation, and is
discussed below in \cref{sec:four-five-int}.  $\mathcal{I}_7$ can be
evaluated using ``bubble iteration'' in terms of ratios of
$\Gamma$-functions, similar to previous loop orders.

\subsection{Analytic evaluations}
\label{sec:seven-int}
Here we state without derivation the analytic expressions needed for
the basis integrals corresponding to
\cref{fig:iter-lower,fig:iter5,fig:o-six,fig:six-o,fig:fourfour,fig:seven}, \ie
\cref{eq:hills-int,eq:four-int,eq:five-int}.
Interested readers can find a more complete discussion in
Ref.~\cite{Edison:2023qvg}, which primarily relies on the integrals
tables from Ref.~\cite{Smirnov:2012gma}.

First, the ratios of $\Gamma$-functions that appear in the recursive
evaluation of the integrals
\begin{align}
  A_{\lambda_1,\lambda_2;d}
  &=
    \frac{
    \Gamma(d/2-\lambda_1)
    \Gamma(d/2-\lambda_2)
    \Gamma(\lambda_1+\lambda_2-d/2)
    }{
    \Gamma(\lambda_1)
    \Gamma(\lambda_2)
    \Gamma(d-\lambda_1-\lambda_2)
    }\,,\\
  B_{\lambda_1,\lambda_2,\lambda_3;d}
  &=
    \frac{
    \Gamma(\lambda_1+\lambda_3 - d/2)
    \Gamma(\lambda_2+\lambda_3-d/2)
    \Gamma(d/2 - \lambda_3)
    \Gamma(\lambda_1+\lambda_2+\lambda_3 - d)
    }{\Gamma(\lambda_1)
    \Gamma(\lambda_2)
    \Gamma(\lambda_1+\lambda_2+2 \lambda_3 -d)
    }\,.
\end{align}
Then the needed integrals
\begin{align}
  \mathcal{I}_{\emptyset}
  &= (-1)(4 \pi)^{-d/2} (-\omega^2)^{d/2-1}
  \label{eq:i0} \,,\\
  \mathcal{I}_4
  &= (4 \pi)^{-3d/2} A_{1,1;d} B_{1,1,2-d/2;d} (-\omega^2)^{3d/2-4}
  \label{eq:i4} \,,\\
  \mathcal{I}_5
  &= (4 \pi)^{-2d} A_{1,1;d} A_{1,2-d/2;d} B_{1,1,3-d;d} (-\omega^2)^{2d-5}
  \label{eq:i5} \,,\\
  \mathcal{I}_6
  &= (4 \pi)^{-5d/2} A_{1,1;d} A_{1,2-d/2;d}A_{1,3-d;d} B_{1,1,4-3d/2;d} (-\omega^2)^{5d/2-6}
  \label{eq:i6} \,,\\
  \mathcal{I}_7
  &= (4 \pi)^{-3d} A_{1,1;d} A_{1,2-d/2;d}A_{1,3-d;d}A_{1,4-3d/2;d} B_{1,1,5-2d;d} (-\omega^2)^{3d-7}
    \label{eq:i7} \,.
\end{align}

\subsection{Numeric evaluation of \cref{fig:ffo,fig:off}}
\label{sec:four-four-int}
The integrals for \cref{fig:ffo,fig:off} factorize into a tadpole
integral and an integral that we refer to as
$\mathcal{I}_{4 \otimes 4}$.  This integral is currently unknown
analytically, but was numerically evaluated \cite{Liu:2022chg} and
reconstructed into known transcendental numbers
\cite{Bailey:1999nv,pslq} in Ref.~\cite{Edison:2023qvg}.  We reproduce
the integral here for completeness.  First, we note that the $\omega$
dependence of the integral can be factorized
\begin{equation}
\mathcal{I}_{4 \otimes 4} = (-\omega^2)^{5d/2-7}\mathcal{I}_{4 \otimes 4}(1)\,.
\end{equation}
We find that the remaining integral evaluates to
\begin{align}
  \mathcal{I}_{4 \otimes 4}(1) = \frac{1}{8(4 \pi)^5}\Bigg[
  & \frac{1}{ \dimreg^2}
    + \frac{16 +5(\log(\pi) - \gamma_E)}{\dimreg} \notag \\
  &+ \big( 184+ \frac{47}{2} \zeta_2 + 80(\log(\pi) -  \gamma_E) + \frac{25}{2}(\log(\pi) - \gamma_E)^2
     \big) \notag \\
  &+\dimreg\Big( 1888+ 408 \zeta_2- \frac{611}{3} \zeta_3 + (\log(\pi) - \gamma_E)\big(920 + \frac{235}{2} \zeta_2\big)\notag \\
  &\quad + 200(\log(\pi) - \gamma_E)^2
  + \frac{125}{6}(\log(\pi) - \gamma_E)^3\Big) \notag \\
  & \dimreg^2\Big(
    18544
    + 5092 \zeta_2
    - \frac{9872}{3} \zeta_3
    + \frac{42193}{40} \zeta_2^2 \notag \\
   &\quad  + (\log(\pi) - \gamma_E)\big(9440 
    + 2040\zeta_2 
    - \frac{3055}{3} \zeta_3\big) \notag \\
    &\quad +  (\log(\pi) - \gamma_E)^2\big(2300
    + \frac{1175}{4} \zeta_2\big) \notag \\
    &\quad + \frac{1000}{3}(\log(\pi) - \gamma_E)^3
    + \frac{625}{24}(\log(\pi) - \gamma_E)^4
      \Big)\notag \\
  &\quad + 255097.168799191\, \dimreg^3 +
      \odimreg{4}\Bigg] \,.
\end{align}
Note the finite precision coefficient of $\dimreg^3$.  We do not yet
have a transcendental representation for this term.  Is this relevant
to the radiated energy computed above?  In the computation of
\cref{eq:t5-action}, we tagged this term and found that it contributes
to the \emph{conservative} piece of the effective action, but not the
\emph{dissipative} piece that we analyze in this paper.

\subsection{Numeric evaluation of \cref{fig:three-two,fig:two-three}}
\label{sec:four-five-int}
The integral needed for \cref{fig:three-two,fig:two-three}, which we
will refer to as $\mathcal{I}_{4 \otimes 5}$, are similar to
$\mathcal{I}_{4 \otimes 4}$.  If we knew an analytic form of
$\mathcal{I}_{4 \otimes 4}$, we could use ``bubble iteration'' to
recursively evaluate both in terms of the same base functions.
Unfortunately, we instead need to perform a numerical evaluation and
PSLQ reconstruction \cite{Bailey:1999nv} of
$\mathcal{I}_{4 \otimes 5}$ as well.  As above, we will factorize the
$\omega$-dependence out of the integral for the numerical evaluation
\begin{equation}
  \mathcal{I}_{4 \otimes 5} = (-\omega^2)^{3d-8} \mathcal{I}_{4 \otimes 5}(1)\,.
\end{equation}
We use AMFlow \cite{Liu:2022chg}
connected to Kira \cite{Klappert:2020nbg} for the numerical
evaluation.  After reconstruction, we find
\begin{align}
  \mathcal{I}_{4 \otimes 5}(1)
  &= \frac{1}{96(4 \pi)^6} \Bigg[
    \frac{11}{\dimreg^2} 
    + \frac{1}{\dimreg}\Big(208 + 66 (\log(\pi) - \gamma_E)\Big) \notag \\
    &\quad + 3(856 + 113 \zeta_2 + 416 (\log(\pi) - \gamma_E) + 66(\log(\pi) - \gamma_E)^2) \notag \\
  &\quad + \dimreg \bigg(26528 + 6672 \zeta_2 - 1658 \zeta_3
    + (\log(\pi) - \gamma_E)\big(15408 + 2034 \zeta_2\big) \notag \\
  &\qquad + 3744 (\log(\pi) - \gamma_E)^2
    + 396(\log(\pi) - \gamma_E)^3
    \bigg) \notag \\
  &\quad \dimreg^2 \bigg(
    251344
    + 86472 \zeta_2
    -  33184 \zeta_3
    + \frac{96837}{4} \zeta_4 \notag \\
  &\qquad + (\log(\pi) - \gamma_E) \big(159168 
    +  40032  \zeta_2
    - 9948  \zeta_3 \big) \notag \\
  &\qquad +(\log(\pi) - \gamma_E)^2\big( 46224 
    + 6102 \zeta_2\big) \notag \\
  &\qquad + 7488 (\log(\pi) - \gamma_E)^3
    + 594 (\log(\pi) - \gamma_E)^4
    \bigg) \notag \\
    &\quad + \dimreg^3\bigg(
      2288256
      + 945312 \zeta_2
      - 438384 \zeta_3
      + 502284 \zeta_4
      - \frac{349086}{5} \zeta_5
      - 60522 \zeta_2 \zeta_3 \notag \\
  &\qquad + (\log(\pi) - \gamma_E)\big(1508064 
        + 518832  \zeta_2
        - 199104  \zeta_3
        + \frac{290511}{2}  \zeta_4 \big) \notag \\
  &\qquad + (\log(\pi) - \gamma_E)^2\big(477504 
      + 120096  \zeta_2
      - 29844  \zeta_3\big) \notag \\
   &\qquad + (\log(\pi) - \gamma_E)^3\big(92448 
      + 12204  \zeta_2\big)
     + 11232 (\log(\pi) - \gamma_E)^4 \notag \\
     &\qquad  + \frac{3564}{5} (\log(\pi) - \gamma_E)^5
      \bigg) + \odimreg{4}
    \Bigg]
\end{align}
Given that the unitarity coefficient of this integral,
\cref{eq:two-three-coef}, contains an $\dimreg^{-3}$ pole, we only
require the value of $\mathcal{I}_{4 \otimes 5}(1)$ through the stated
\odimreg{3} in order to extract the finite contribution to the
effective action and energy loss.

\bibliographystyle{JHEP}
\bibliography{gwbibtex}

\end{document}